\documentclass[aps,prl,twocolumn,groupedaddress]{revtex4}

\usepackage[]{graphicx}
\usepackage{amsmath}       
\usepackage{amssymb}       
\usepackage{url}

\begin{document}

\begin{widetext}
\noindent\textbf{Preprint of:}\\
Katrina Y. T. Seet, Robert Vogel, Timo A. Nieminen,
Gregor Kn\"{o}ner, Halina Rubinsztein-Dunlop,\\
Matt Trau, and Andrei V. Zvyagin\\
``Refractometry of organosilica microspheres''\\
\textit{Applied Optics} \textbf{46}(9), 1554--1561 (2007)\\
\textcopyright 2001 Optical Society of America, Inc.
\end{widetext}

\title{Refractometry of organosilica microspheres }

\author{Katrina Y. T. Seet}
\affiliation{Center for Biophotonics and Laser Science, School of Physical
Sciences, The University of Queensland\\ \emph{and}\\
School of Information Technology and Electrical Engineering,
The University of Queensland}
\email{seet@physics.uq.edu.au}

\author{Robert Vogel}
\affiliation{Nanotechnology and Biomaterials Centre,
The University of Queensland}

\author{Timo A. Nieminen}
\author{Gregor Kn\"{o}ner}
\author{Halina Rubinsztein-Dunlop}
\affiliation{Center for Biophotonics and Laser Science,
School of Physical Sciences, The University of Queensland}

\author{Matt Trau}
\affiliation{Nanotechnology and Biomaterials Centre,
The University of Queensland, Brisbane, 4072, Australia}

\author{Andrei V Zvyagin}
\affiliation{Center~for Biophotonics and Laser Science, School of Physical
Sciences, The University of Queensland\\ \emph{and}\\
School of Information Technology and Electrical Engineering,
The University of Queensland}

\begin{abstract}
The refractive index of novel organosilica (nano/micro)material is
determined using two methods.  The first method is based on analysis of
optical extinction efficiency of organosilica beads versus wavelength,
which is obtained by a standard laboratory spectrometer.  The second
method relies on the measurable trapping potential of these beads in the
focused light beam (laser tweezers).  Polystyrene beads were used to
test these methods, and the determined dispersion curves of refractive
index values have been found accurate.  The refractive index of
organosilica beads has been determined to range from 1.60--1.51 over
the wavelength range of 300--1100 nm.

\end{abstract}

\maketitle 

\section{Introduction}

Properties of (nano/micro)materials are shaped up by the bulk and
structural properties of their building blocks, with an example being
refractive index that determines the material's optical characteristics. 
These optical properties need to be precisely controlled in various
applications, such as sizing of organosilica (nano/micro)beads in the
flow cytometry biological assaying. By manipulating the composition of
materials on the nanoscale, new structural properties have been
produced.  For reliable production of desirable (nano/micro)materials
both bulk and structural properties need to be controlled.  Since bulk
properties of the material can vary as a result of nanotechnological
synthesis, it is important to monitor them \textit{in situ}.  In
particular, silica-derived novel material, termed organosilica, have
been synthesized in the form of (nano/micro)spheres, with a range of
useful qualities that have been
reported~\cite{Miller200538,Miller200521}.
Organosilica surface activation have appeared to be
straightforward to bind oligonucleotide and peptide sequences for
optical encoding and high-throughput screening via flow
cytometry~\cite{Miller200538,Miller200521}. The optical encoding method,
however, demands tight margins on organosilica microspheres size and
refractive index to ensure reliable readout and correct interpretation
of their scattering intensities in the context of flow cytometry. 
However, it has been noticed that the refractive index of organosilica
microspheres is highly dependent on synthesis
route~\cite{vanBlaaderen1993156} calling for a reliable refractometry
method amenable for in situ implementation.  Since the organosilica
synthesis yields spherical particles ranging from 50 nm to 6 $\mu$m, a
number of existing refractometry methods applicable for bulk materials
are no longer suitable for refractometry of this material.  Among the
suitable methods, an index matching method has been reported in a number
of particle refractometry
applications~\cite{GarciaSantamaria200218,vanBlaaderen1993156}.
In brief, particles in the solution do not
scatter, i.e. rendering the sample clear, if refractive indices of the
particles and solution are matched.  Therefore, the solution refractive
index read out at the sample maximum transmittivity provides a measure
of the refractive index of the particle ensemble.  Despite its
conceptual simplicity, the method is tedious, requiring preparation of
the new solution of predetermined refractive index per each experimental
data point.  The availability of suitable solutions with high refractive
indices is usually limited.

The majority of existing refractometry methods for particles rely on the
acquisition of optical scatter data, followed by their fitting to the
theoretical model from which one can infer the particle refractive
index, or its size, or both.  For example, the determination of
refractive index of an ensemble of microspheres have been demonstrated
using holographic method of recording of the optical
scatter~\cite{Alexandrov200328}.  The intensity of optical scatter has
been parameterized in the Rayleigh scattering model to perform sizing of
nanodiamonds~\cite{Colpin200631}.  Differential microscopy measurements
of the optical scatter have also permitted detection of the minute cell
organelle swelling, i.e. change in the scatterer's size, as a result of
induced apoptosis~\cite{Boustany200126}.
 
Scattering spectroscopy provides another approach in the determination
of particle refractive index in which characteristic particle optical
extinction efficiency versus wavelength is
determined~\cite{Chen2003228,Chylek198322,Guimaraes199289,%
Scholz199883,Liu200530,Ma200348}.
Usually, the acquired spectrum of monodisperse microsphere samples is
fitted to a theoretical model with the refractive index and microsphere
size as parameters.  Using this method, the refractive index for a
specified wavelength range, i.e. refractive index dispersion, can be
determined.  The application of the scattering spectroscopy technique to
refractometry of cells, bacteria, and spores has been also
reported~\cite{Alupoaei200419,Callahan200337,%
GarciaSantamaria200218,Mattley200071}.

In this paper, we demonstrate the application of two refractometry
methods to determine the refractive index of novel organosilica
microspheres.  Firstly, we report application of the refined scattering
spectroscopy method, where we use a commercial laboratory Lamda40
spectrophotometer (Perkin Elmer) to measure the refractive index and
dispersion of organosilica, following well-characterized polystyrene
microsphere refractometry.  Secondly, we demonstrate application of the
recently reported method for measurement of the refractive index for
single microparticles based on optical tweezers~\cite{Knoner200697}.  The
two methods are compared, and a very good agreement is achieved.

\section{Experimental Method}

\subsection{Materials}

Organosilica spheres were produced using a two-step
process~\cite{Miller200521}.  This process involved the hydrolysis and
subsequent condensation of thiol-based silica precursor.  Aqueous
solutions of 5.94 $\mu$m organosilica microspheres, with the addition of
sodium dodecyl sulphate for the prevention of agglomeration, were
prepared for spectroscopy measurements.  Polystyrene microspheres
(ProSciTech, 5.0\% w/v in water) of mean diameters of 2.01, 2.09 and
5.26 $\mu$m, with a coefficient of variance (CV, defined as the standard
deviation divided by the mean), of less than 2.4\%, were diluted with
MilliQ water.  For the laser tweezers refractometry, 2.09 $\mu$m
polystyrene and 5.16 $\mu$m organosilica microspheres were suspended in
MilliQ water.  Samples were prepared by placing a drop of the suspension
between the microscope slide and cover slip and sealing the chamber with
high viscosity silicon grease.

\subsection{Methods}

\subsubsection{Scanning Electron Microscopy (SEM)}

In order to characterize the size distribution of the synthesized
organosilica microspheres, SEM was used.  Samples were platinum coated
and imaged using SEM (JEOL JSM 6400F) with accelerating voltage of
5--10 kV.  The size distribution of the microspheres was obtained by
processing the acquired images followed by statistical analysis.

\subsubsection{Scattering Spectroscopy} \label{Sub:ScatSpec}

In scattering spectroscopy, spherical particle extinction efficiency
verses illumination wavelength (hereafter referred to as the scattering
spectrum) was recorded.  Recorded data were fitted to the theoretical
model, where fitting parameters represented refractive index and size of
the particle.  To determine the best fit, the  value, defined as the sum
of the difference squared $\sum\left(I_{exp} - I_{theor}\right)^2$, was
minimized.  Although this method is simple and effective, the final best
fit can be significantly unalike.  This is due to the recording
configuration, which should be a simple plane wave incident on a single
spherical particle that cast a shadow on a small-size photodetector
situated at a distance.  The attenuation of the incident optical flux
along the axis of incidence would be entirely due to light scattering by
the particle, or in other words, the particle scattered light away from
the photodetector.  Under typical experimental conditions, however,
illumination from an extended light source could not be regarded as the
plane wave illumination.  In addition, instead of a single particle, one
has to account for an ensemble of particles, and a finite photodetector
size.

\begin{figure} 
\centerline{\includegraphics[width=\columnwidth]{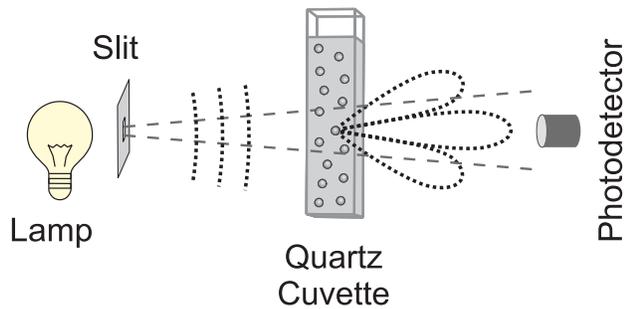}}
\caption{A lamp produces a wave with divergence of $0.4^{\circ}$.  This
is incident on a quartz cuvette which is filled with an aqueous
suspension of microspheres.  A silicon photodetector measures
transmission intensity for wavelengths 300--1100 nm.}
\label{FigSpectrometerSetup}
\end{figure}

In this paper, we employed a commercial Lamda40 spectrophotometer
(Perkin Elmer) to acquire sample extinction spectra.  The sample was an
aqueous suspension of polystyrene or organosilica microspheres in a
square-profile (10 mm $\times$ 10 mm) quartz cuvette.  The concentration
of the suspension was varied (average $1.5 \pm 0.6$ g/L) to ensure
operation in the single-scattering regime without compromising the
signal-to-noise ratio.  Prior to spectroscopy, every suspension was
ultrasonicated for 3 minutes to ensure uniform suspension of
microspheres.  To verify suspension uniformity, spectra measurements
were repeated immediately after the first recording. Both spectra were
identical, demonstrating no particle drift occurred during measurements.
The cuvette was illuminated by a light source that produced a beam spot
of approximately $1.0 \times 7.5~\textrm{mm}^{2}$ on the front facet of
the cuvette, and was characterized by an angular divergence of
$0.4^{\circ}$.  A halogen lamp was used as the light source, in the
visible and near-infrared spectral range, whereas a deuterium lamp was
used in the ultraviolet spectral range.  The illumination light source
wavelength was selected by means of three spectrally adjacent tunable
diffraction gratings whose operation wavelength range spanned 300 - 1100
nm.  The detector was a silicon photodetector with an area of
approximately 49 mm$^{2}$ calibrated for the wavelength range used. The
source to detector configuration is shown in Fig.
\ref{FigSpectrometerSetup}.

In addition, a cuvette with pure water was placed in the reference path
of the spectrometer. Optical transmission spectra of the sample and
reference were recorded over the wavelength range of 300 - 1100 nm, and
each reference spectrum was subtracted from the sample spectrum to
account for reflections, aberrations and the intrinsic absorption
profile of the solvent and cuvette material.  Scans were performed at a
rate of 480 nm min$^{-1}$ with a spectral resolution of 2 nm.  Recorded
data were digitized and transferred to a personal computer for further
processing, archiving, and display.  The recorded data were fitted to
the theoretical scattering spectra, as detailed in the modelling method
in Section \ref{sec:modelmeths}.

\subsubsection{Laser Tweezers Refractometry}

\begin{figure}[bth]
\centerline{\includegraphics[width=\columnwidth]{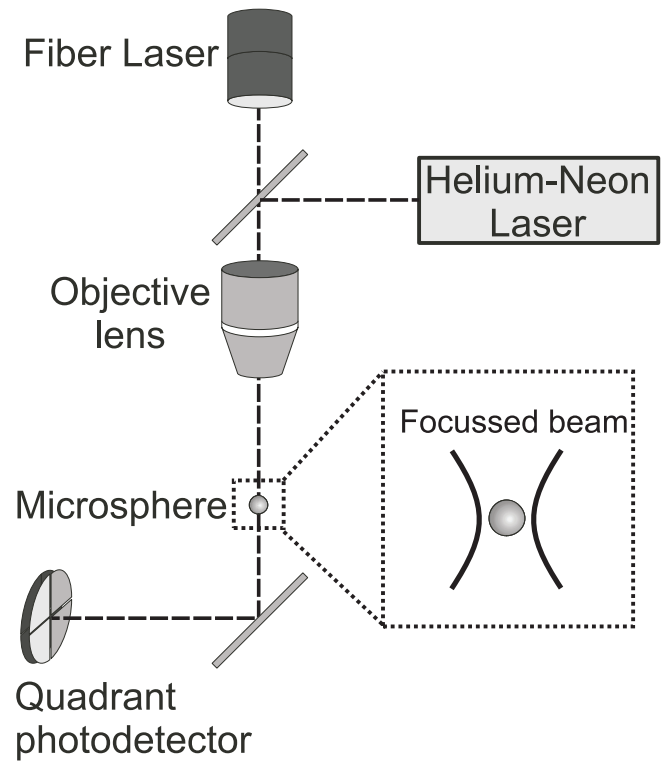}}
\caption{Schematic diagram of the optical tweezers refractometry for
refractive index measurement of individual microspheres.  A microsphere
was trapped by a tightly focused laser beam.  The microsphere
displacement from the centre of the trap was detected by illuminating it
with a helium-neon laser and detecting the signal on the quadrant
photodetector.  The inset shows the microsphere trapped at the centre of
the focused fiber laser beam.}
\label{FigTweezerSetup}
\end{figure}

The second refractometry method is based on laser
tweezers~\cite{Ashkin198611,Knoner200697}.  Here, we measure the particle
trapping potential created by a high-power focused laser beam, as shown
in Fig. \ref{FigTweezerSetup}.  For a particle displaced by a small
distance, $\Delta l$, in transverse direction, the acting force,
F$_\alpha$, is measured. From this, the trap stiffness expressed in
terms of the trap stiffness parameter, $\alpha$, is calculated by
$F_\alpha/\Delta l$ (measured in piconewtons per nanometre, pN/nm). This
stiffness was calculated for a particle of specified size for a range of
refractive indices of the particle.  From the theoretical analysis of
measured results, as described in Section \ref{sec:opttweezers} we
obtained the refractive index of the microsphere.  In order to create
the optical particle trap, the beam from a continuous-wave fiber laser
(power 1 watt, wavelength $\lambda=1070$\,nm) was focused to a
diffraction-limited spot using an objective lens with high numerical
aperture of 1.3.  In order to detect the particle position in the beam,
light from a helium-neon laser scattered by the particle was detected on
a quadrant photodetector.  Alternatively, the fiber laser was used both
for trapping and tracking of the particle position.  In both cases, the
particle displacement from the trap centre generated a misbalanced
photoelectrical signal on the photodetector quantifiable by means of the
electronic signal processing~\cite{Knoner200572,Lang200283}.

\section{Modelling Methods} \label{sec:modelmeths}

\subsection{Scattering Spectroscopy}

\begin{figure}[bht]
\centering{\includegraphics[width=\columnwidth]{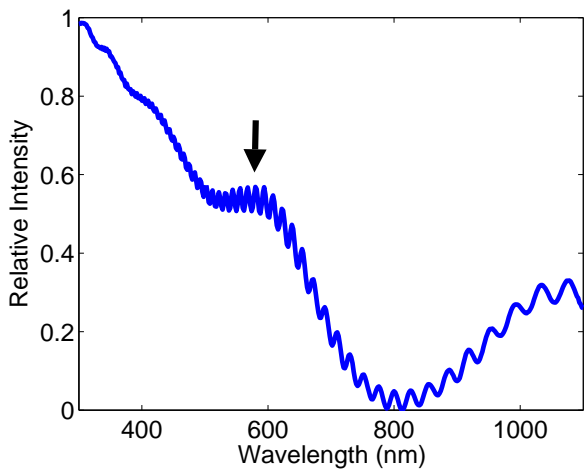}}
\caption{Typical scattering spectra of microsphere suspensions in
solution. (a) Three distinct features can be observed: ripples, local
maxima, and a sloping curve. (b) Ripple feature of the scattering
spectra, extracted by high-pass filtering.}
\label{FigExampleSpec}
\end{figure}

A typical scattering spectrum [Fig. \ref{FigExampleSpec}(a)] exhibits
three major features: ripples, local maxima, and a sloping curve.  The
ripple represents a small-contrast high-frequency sinusoidal signal
superimposed on the large background signal.  An example of the filtered
ripple structure signal is shown in Fig. \ref{FigExampleSpec}(b).  This
ripple signal is the result of interference between the incident and
light internally reflected within the particle.  It can provide an
accurate measure of the product of microsphere diameter $D$ and
refractive index $n_{sp}$.  It is important to note that the ripple
signal is detectable only if the sample is monodisperse and the size
parameter, $x$, is much greater than 1.  Otherwise, the ripple structure
of the signal is smeared and ultimately
undetectable~\cite{Chylek198322,Chylek19896}.
The light source and detector configurations have little
effect on the ripple structure of the scattering spectrum.

A local maxima, in Fig. \ref{FigExampleSpec}(a). can be explained by the
Lorenz-Mie scattering theory.  In brief, it is due to interference of
the incident and diffracted light at the particle-medium interface.  A
phase delay acquired by the diffracted light depends on the particle
diameter and its relative refractive index $n_{sp}/n_{m}$ , where
$n_{m}$ is the refractive index of the medium.  By thin film
approximation and consideration of the optical path length difference,
the spectral separation between the maxima of the scattering spectrum
curve can be determined~\cite{Scholz199883}:

\begin{equation}
\Delta\lambda=\left( \lambda_{(i+1)}-\lambda_i \right)\approx \frac{\lambda^2_i}{D\left(n_{sp}-n_m\right)}
\label{EqGlobalPeaks}
\end{equation}

where $i$ and $i+1$ stand for the adjacent peaks wavelength indices.  We
note that the fidelity of this relationship is affected by the
source-detector configuration.  However, in the context of our work, it
provides a good estimation of particle refractive index.

The sloping curve representing the scattering spectrum is shown in Fig.
\ref{FigExampleSpec}(a) and is tailing towards the long wavelength side. 
This curve also stems from the Lorenz--Mie scattering theory.  In brief,
the optical scatter of a microsphere takes the form of scattering lobes. 
The larger the size parameter, the more pronounced and narrow its
forward-scattering lobe.  Alternatively, the shorter the wavelength of
light incident on a sample of microspheres, the less power is scattered
away from the photodetector, hence the greater intensity is detected. 
This consideration qualitatively explains the sloping trend of the
recorded spectra, although accurate theoretical modelling is needed for
precise fitting of experimental data to the theory, and is beyond the
scope of this paper.

In order to obtain a dispersion curve for the refractive index, a
Lorenz--Mie scattering model was used to calculate the extinction
cross-section of monodisperse microspheres suspended in a homogeneous
medium.  These calculations assume an incident plane wave and a
infinitesimally small-size far-field detector i.e. collection angle of
zero~\cite{vandeHulst1957}.  Refractive index dependence on the
wavelength was described by the Cauchy dispersion relation.  Its general
form is given by

\begin{equation}
n\left(\lambda\right)=A +\frac{B}{\lambda^2}+\frac{C}{\lambda^4}
\label{EqCauchy}
\end{equation}

where the Cauchy coefficients A, B ($\mu$m$^2$) and C ($\mu$m$^4$) are
unique to the material~\cite{Jenkins1976}.  Using an unconstrained
nonlinear optimization method implemented in Matlab, particle size and a
dispersion curve of refractive indices over a wavelength range of
300--1100 nm were determined.

\subsection{Optical tweezers refractometry} \label{sec:opttweezers}

Optical tweezers refer to a technique of trapping a single microsphere
immersed in fluid at the focus of a laser beam.  It is well known that
particles in fluid experience random Brownian motion.  For a trapped
microsphere, Brownian motion is suppressed, where the suppression
efficiency is a monotonic function of the optical trap stiffness,
$\alpha$, which in turn depends on the microsphere size and refractive
index.  Therefore, the analysis of the Brownian motion amplitude versus
the controllable stiffness provides a basis for the determination of the
particle refractive index.  The Brownian motion amplitude is determined
by detecting the microsphere position fluctuation in the trap, which is
analysed in terms of its power spectrum amplitude.  The roll-off
frequency, $f_\circ$, of this spectrum provides a reliable measure of
the trap stiffness via the following relationship~\cite{Knoner200697}:

\begin{equation}
f_\circ=\frac{\alpha}{2\pi\beta}
\end{equation}

where $\beta$ is the drag coefficient well tabulated for various fluids. 
$\alpha$ relates to $x$ and $n_{sp}$.  This dependence has been analysed
in the framework of the generalized Lorenz-Mie theory and detailed
elsewhere~\cite{Knoner200697,Nieminen20045514,Nieminen200379}.
Experimental measurement of $\alpha$ and its comparison with the
theoretical curve of $\alpha$ versus $n_{sp}$ at a fixed $x$, yields the
refractive index of the particle.

\section{Results and Discussion}

\begin{figure}[htb]
\centerline{\includegraphics[width=\columnwidth]{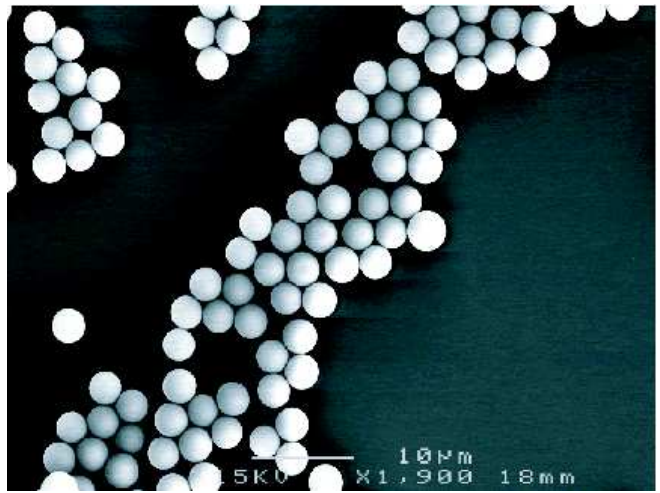}}
\caption{SEM image of monodisperse organosilica microspheres}
\label{FigSEM}
\end{figure}

A low-dispersion particle sample is required for precise determination
of refractive indices of microspheres by means of scattering
spectroscopy.  Fig.~\ref{FigSEM} shows a typical ensemble of
organosilica microspheres imaged by SEM.  The CV of organosilica
microspheres was determined to be less than 10\%, suitable for the
scattering spectroscopy measurements.

\begin{figure*}[p]
\centering
\begin{tabular}{cc}
\large{(a)} & \large{(a$'$)} \\
\includegraphics[width=0.48\textwidth]{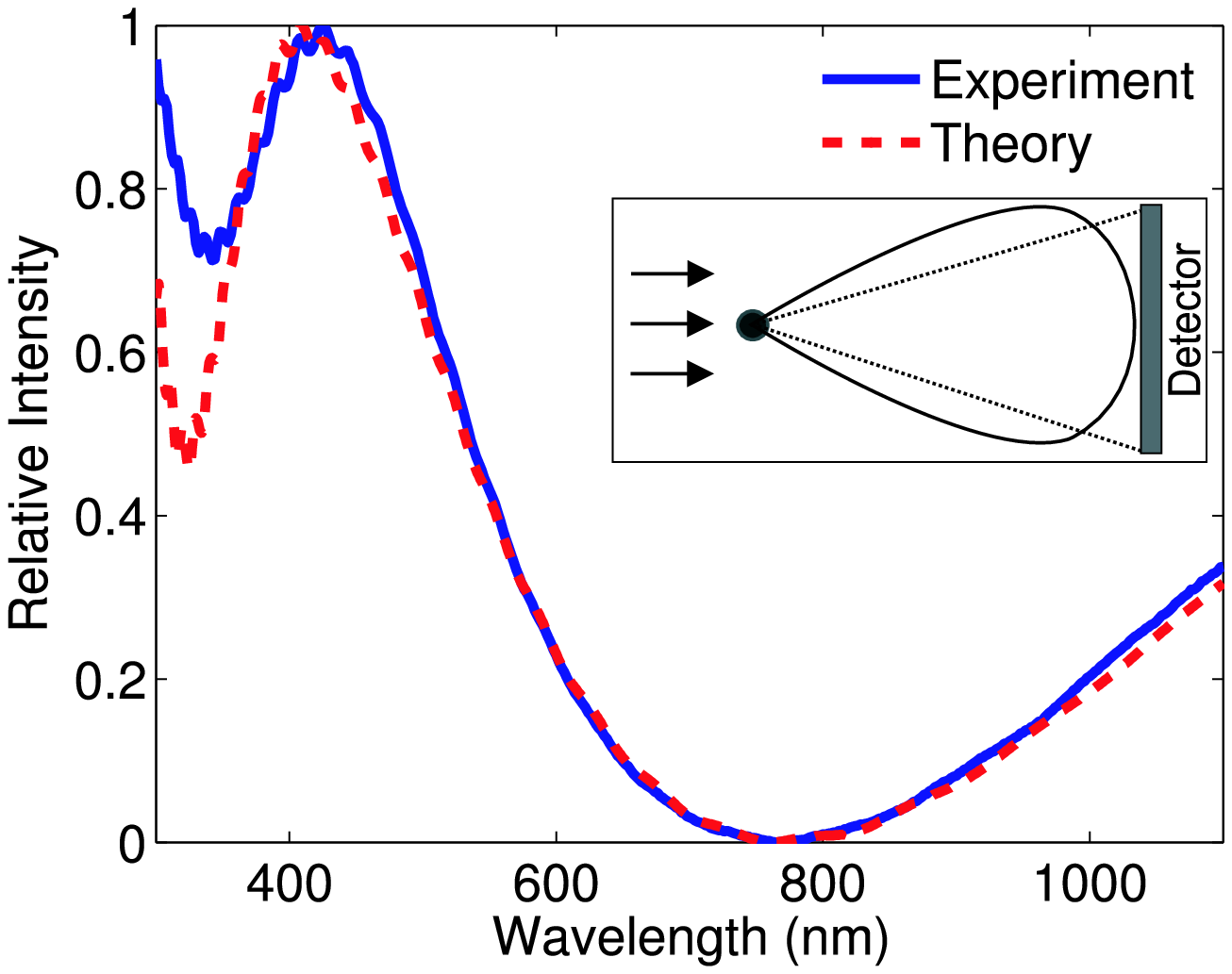}
& \includegraphics[width=0.48\textwidth]{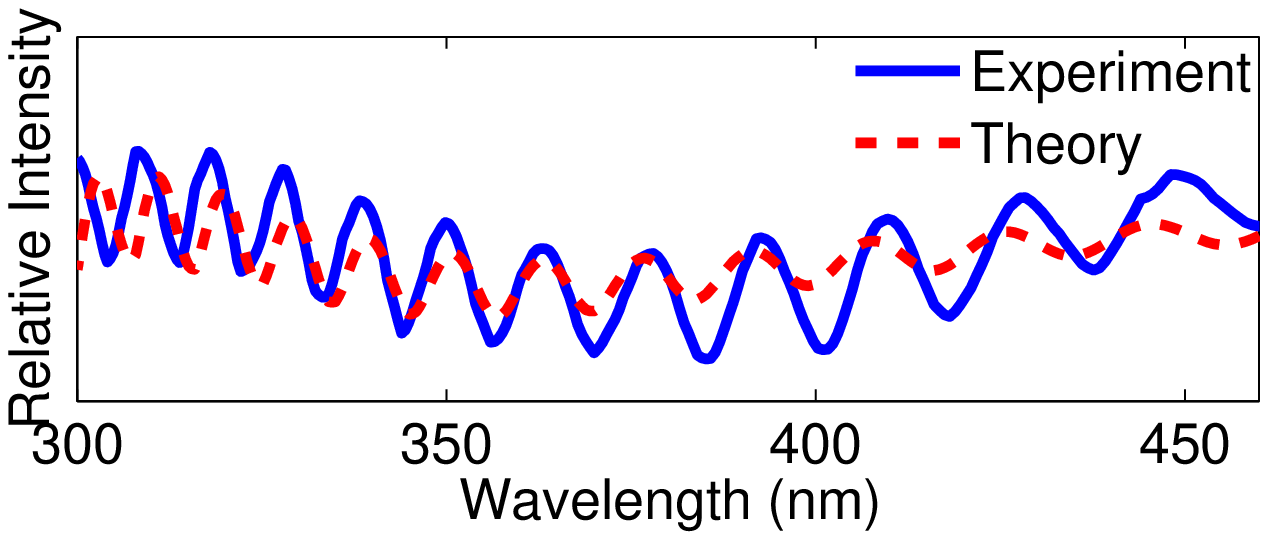}\\
\large{(b)} & \large{(b$'$)} \\
\includegraphics[width=0.48\textwidth]{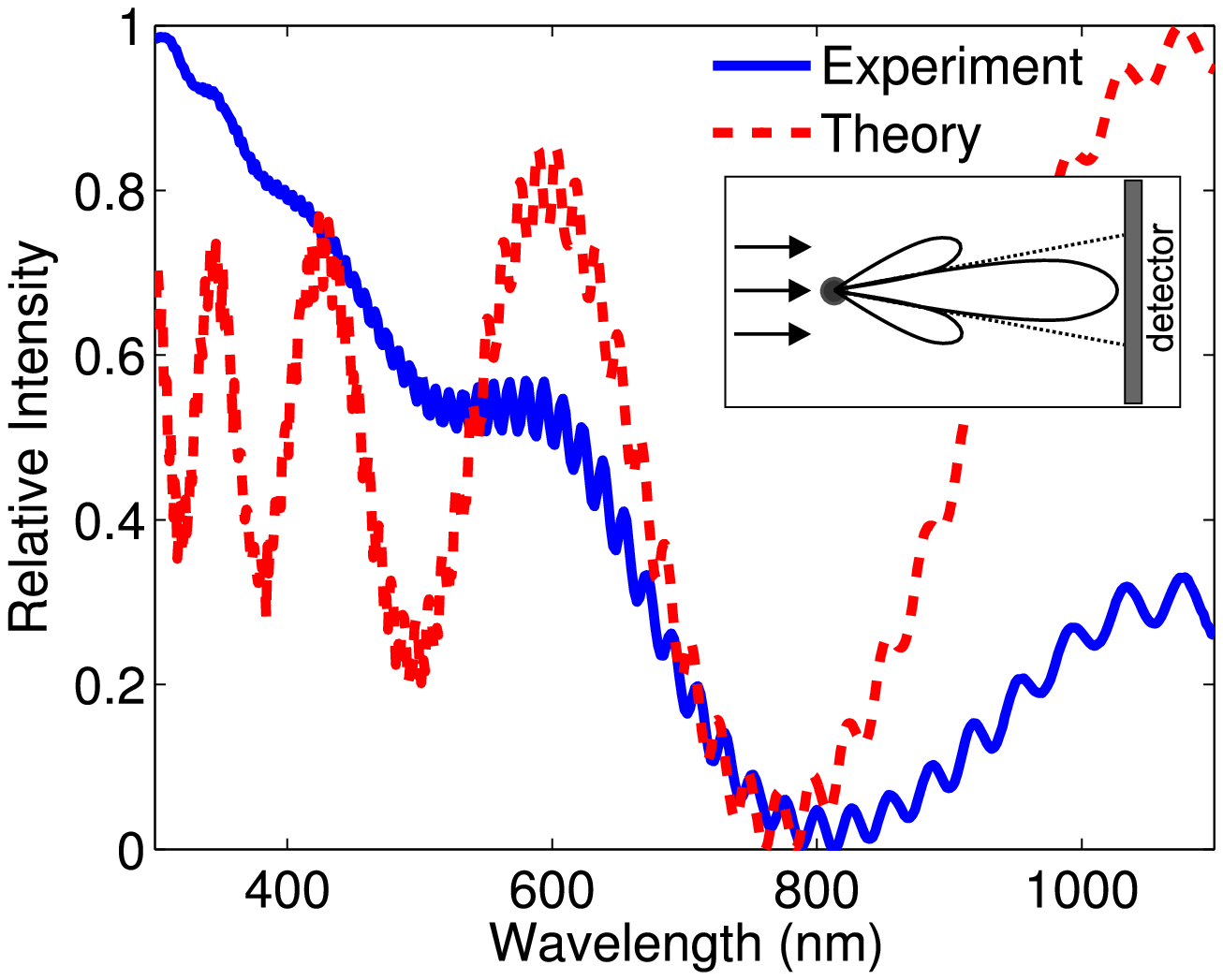}
& \includegraphics[width=0.48\textwidth]{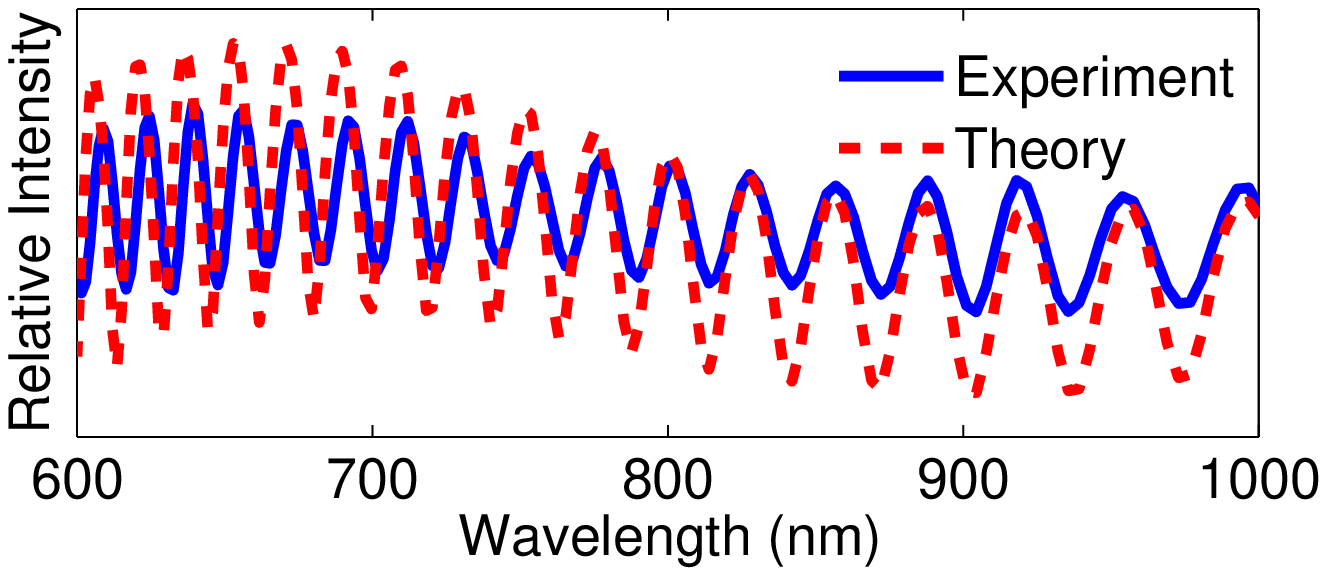}\\
\large{(c)} & \large{(c$'$)} \\
\includegraphics[width=0.48\textwidth]{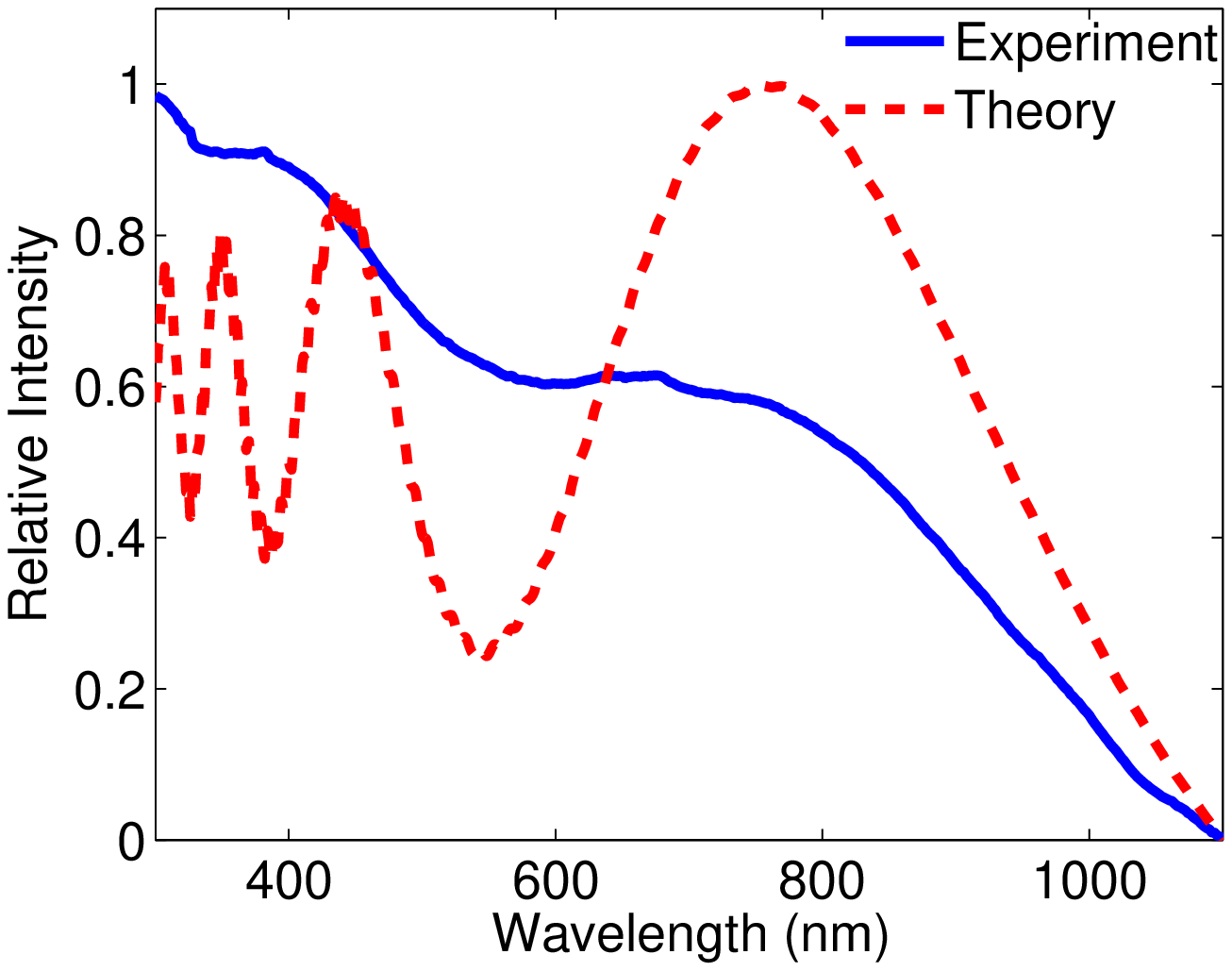}
& \includegraphics[width=0.48\textwidth]{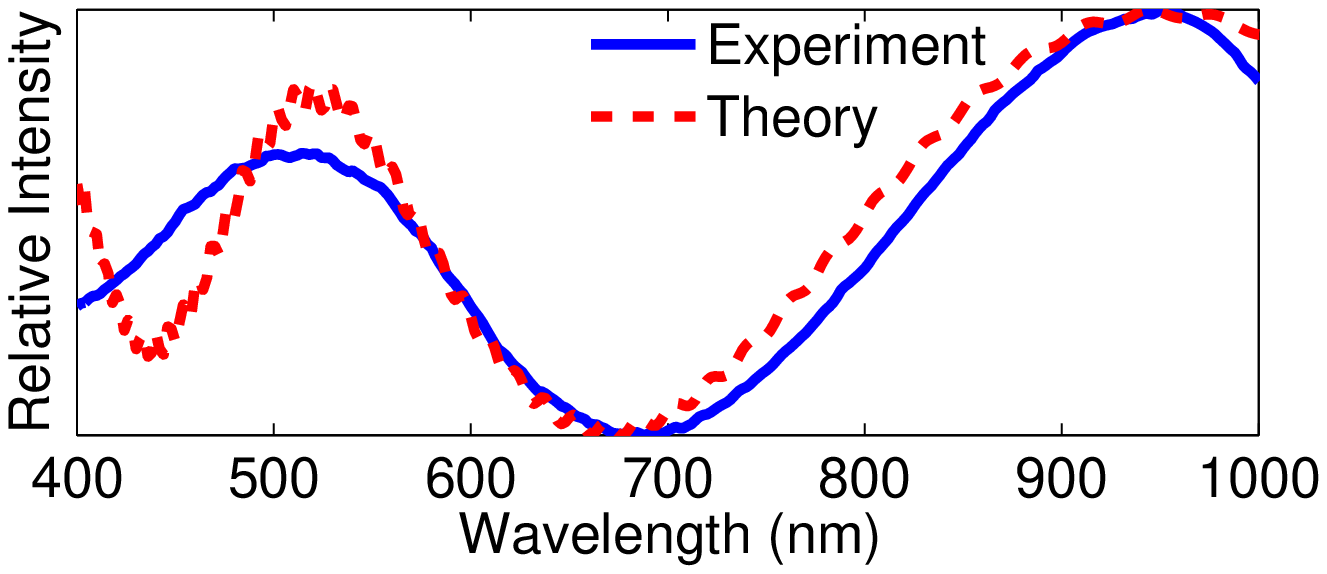}\\
\end{tabular}
\caption{Recorded and calculated spectra of microsphere suspension of
(a) 2.01 $\mu$m and (b) 5.26 $\mu$m, polystyrene and (c) 5.94 $\mu$m
organosilica microspheres in water.  High-pass filtered spectra of
(a$'$) 2.01 $\mu$m and (b$'$) 5.26 $\mu$m polystyrene, and (c$'$) 5.94
$\mu$m organosilica microsphere suspensions.}
\label{FigScatSpec}
\end{figure*}

\begin{figure}
\centering
\begin{tabular}{c}
\large{(a)}\\
\includegraphics[width=\columnwidth]{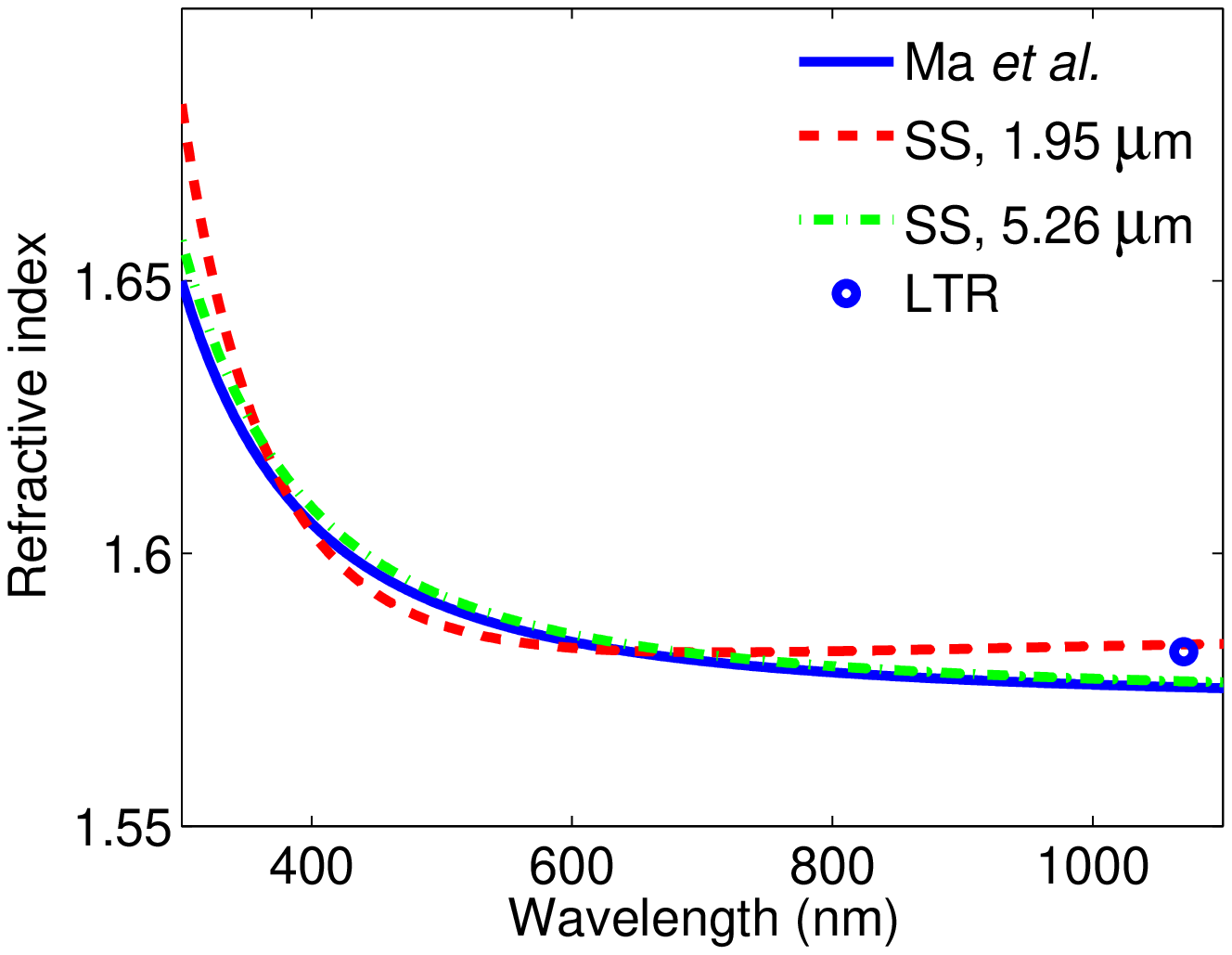}\\
\large{(b)}\\
\includegraphics[width=\columnwidth]{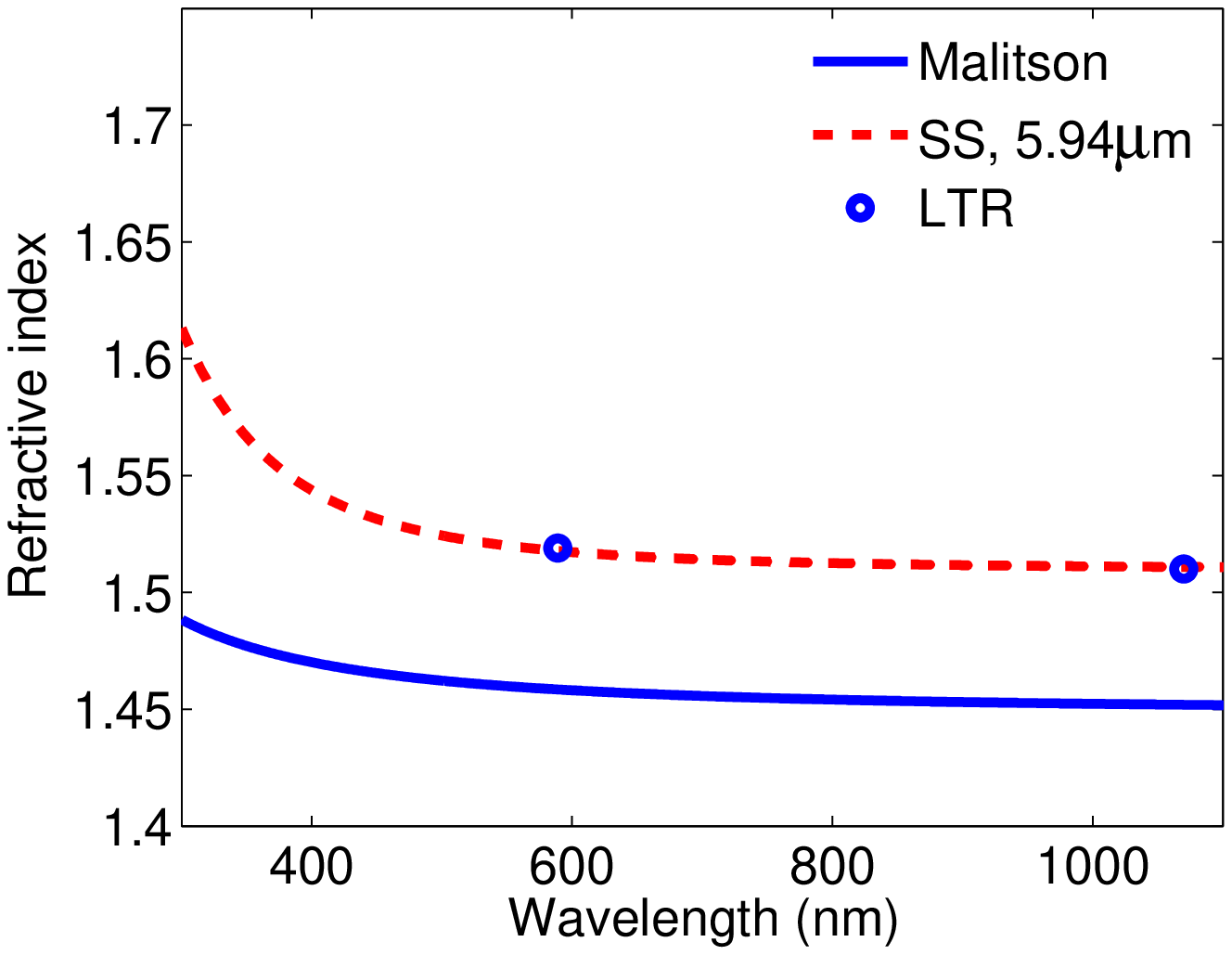}\\
\end{tabular}
\caption{Refractive index dispersion of (a) polystyrene and (b)
organosilica microspheres by scattering spectroscopy (SS) and laser
tweezers refractometry (LTR).  ``Ma et al.'' denotes   for polystyrene,
as given by Ma et al. (2003)~\cite{Ma200348}; ``Malitson'' denotes
refractive index dispersion for fused silica, as given by Malitson
(1965)~\cite{Malitson196555}.}

\label{FigDispersion}
\end{figure}

Organosilica microspheres with well controlled diameters could be
produced in the size range of 1--6 $\mu$m.  In Fig.
\ref{FigScatSpec}(a), (b), and (c), recorded and calculated spectra for
2.01 and 5.26 $\mu$m polystyrene and 5.94 $\mu$m organosilica
microsphere samples, respectively, are presented.  Large size
organosilica microspheres were chosen to ensure the experimental spectra
exhibited at least two maxima.  In the case of the monodisperse 2.01
$\mu$m polystyrene microsphere aqueous suspension sample, the recorded
scattering spectrum was accurately fitted by the calculations using the
Lorenz-Mie theory.  The calculations of the dispersion of polystyrene
refractive index are presented in Fig. \ref{FigDispersion}(a).  The
theoretical fitting curve was parameterized in terms of the Cauchy
coefficients (presented in Table \ref{TabCauchy}), which shows very good
agreement with those reported in the literature.  We believe that the
good agreement between the experimental and theoretical data was due to
the relatively small sphere size, which forward-scattered incident light
over wide angles.  The evaluation of the 2.01 $\mu$m polystyrene sphere
scattering showed that the forward-scattering lobe overfilled the area
of the photodetector for $\lambda \le$ 390 nm, as illustrated in the
Fig. \ref{FigDispersion}(a) inset.  This qualitative model also explains
the discrepancy between theoretical and experimental data at $\lambda
\le$ 390 nm, where the next side-lobe incident on the detector
interferes constructively with the original incident plane wave, unlike
the forward-scattering lobe which interferes destructively (which is why
extinction in the forward direction occurs).  Therefore, the extinction
is reduced in comparison with that calculated in the small photodetector
approximation.  Alternatively, one can view this in terms of the size
parameter $x$.  In Fig \ref{FigScatSpec}(a), the introduction of the
second side-lobe can be observed at an $x$ value of 17.  The effect of
multiple side-lobe interference is much more pronounced in the case of
the scattering spectrum of 5.26 $\mu$m polystyrene microspheres, shown
in Fig. \ref{FigScatSpec}(b), and 5.94 $\mu$m organosilica microspheres,
shown in Fig. \ref{FigScatSpec}(c).  Here, the maximum $x$ values are 15
and 17.  The insets of Fig. \ref{FigScatSpec}(a) and (b) illustrate the
detection scenarios for 2.01 and 5.26 $\mu$m polystyrene sphere aqueous
suspension, respectively.  Obviously, reducing the detector size can
avoid the collection of the side-scattering lobe.  However, commercially
available spectrometers have relatively large size detectors, which are
designed to maximize detection sensitivity.

For an accurate theoretical model, which will result in precise fitting
of measured data, a finite detector must be accounted for.  However,
this will require more complex Mie theory calculations and enormous
computing power.  Therefore to obtain maintain the simplicity of this
method, we employ high pass filtering to isolate the ripple signal from
the sloping curve of the recorded spectra.  The results are presented in
Fig. \ref{FigScatSpec}(a$'$) and (b$'$) for 2.01 and 5.26 $\mu$m
polystyrene microspheres, respectively.  For the 5.26 $\mu$m polystyrene
sample, we used the ripple signal to determine the refractive index
dispersion relationship, provided an independent measurement of the
particle size could be made.  For the organosilica microspheres, the
ripple structure was not observable due to relatively large CV value. 
Therefore, we used the local maxima to determine their refractive index. 
The maximum was isolated by a high-pass filter to suppress the sloping
feature of the recorded spectrum.  The resultant spectrum is shown in
Fig. \ref{FigExampleSpec}(c$'$), and displays a reasonable agreement
with the theoretical scatter spectrum.

\begin{table*}
\begin{center}\caption{Cauchy coefficients for refractive index
dispersion relation (Eq. \ref{EqCauchy})}
\label{TabCauchy}
\begin{tabular}{| l | r | r | r | r |}
\hline
Microsphere & A & B & C & $\lambda$\\
 & & ($\times$10$^{-3} \mu$m$^{-2}$) &	($\times$10$^{-4} \mu$m$^{-4}$)& (nm) \\ \hline
Ma et al. polystyrene\cite{Ma200348}& 1.5725 & 3.1080	& 3.4779	& 390-1310 \\
2.01 $\mu$m polystryene	& 1.5865	& -4.700	& 12.0	& 390-1100 \\
5.26 $\mu$m polystyrene &	1.5736 &	3.0499 &	4.0497 &	390-1100 \\
Fused Silica\cite{Malitson196555} & 1.4492 &	3.1200	& 0.3670	& 200-2000 \\
Organosilica	& 1.5169	& 0.4231	& 6.6304	& 430-1100 \\ \hline
\end{tabular}
\end{center}
\end{table*}

Multiple scattering, other than in the forward direction, was not
accounted for in our modelling method.  The effect of multiple
scattering has been shown, in literature, to increase with optical depth
(concentration) and gradually smoothes out the maxima and
minima~\cite{Cohen197514}.  The spectra of solutions with increasing
concentrations were recorded, and did not exhibit the effects of
multiple scattering, which were observed in the work of Cohen (1975).

\begin{figure} 
\centerline{\includegraphics[width=\columnwidth]{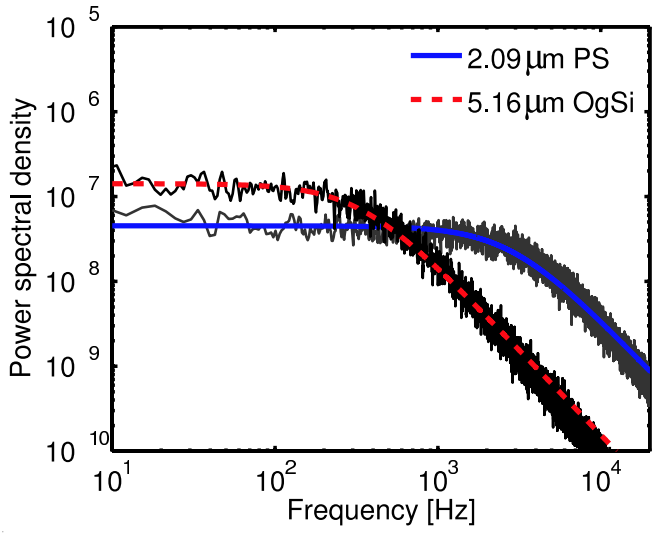}}
\label{FigPowerSpec}
\caption{Plot of power spectral amplitude densities of 2.09 $\mu$m
polystyrene (PS) and 5.16 $\mu$m organosilica microspheres (ogSI) versus
frequency of particle Brownian motion in the optical tweezers trap.  The
roll-off frequency ($f_o$) was used to determine the trap stiffness.}
\end{figure}

\begin{figure} 
\begin{tabular}{c}
\centerline{\includegraphics[width=\columnwidth]{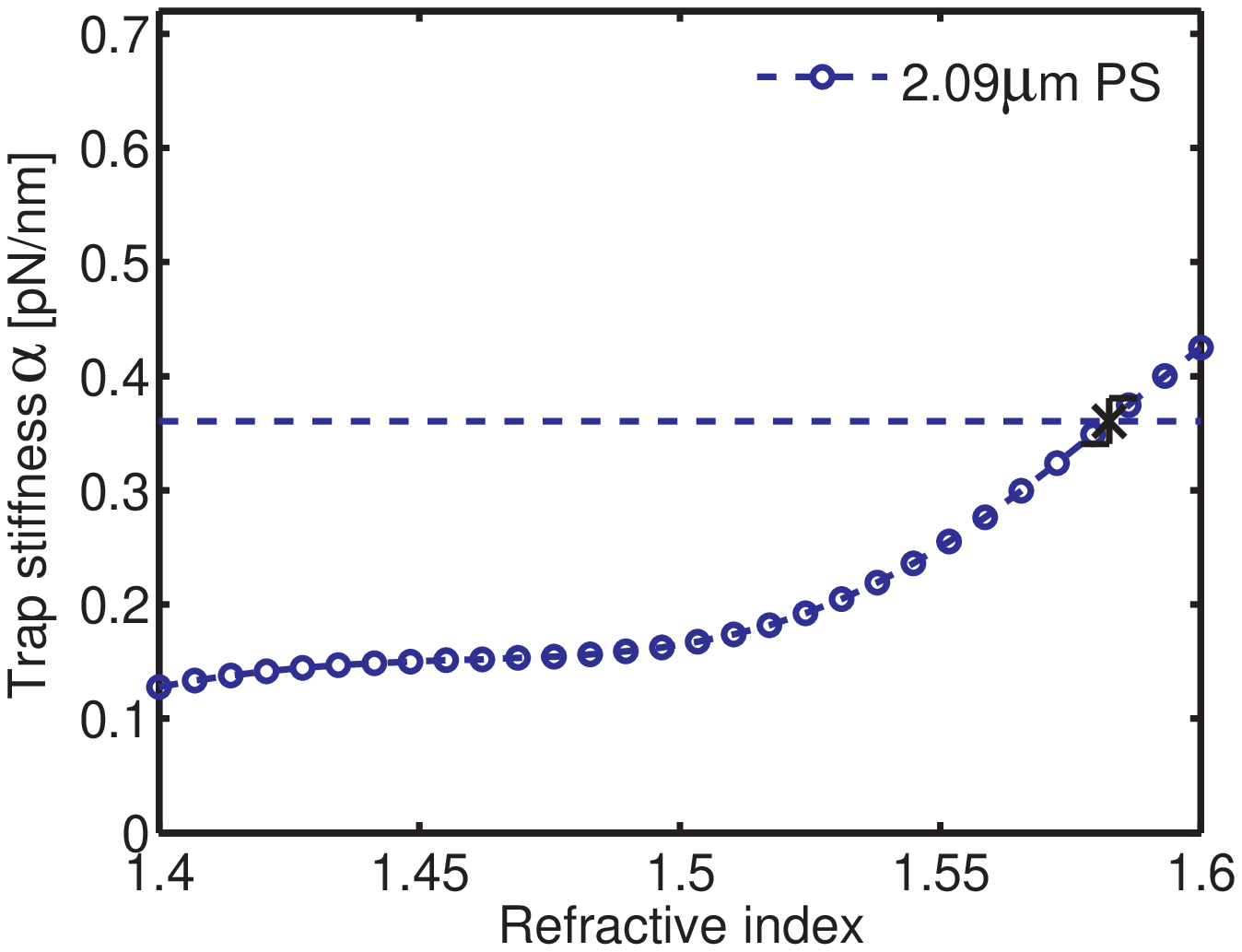}}\\
\centerline{\includegraphics[width=\columnwidth]{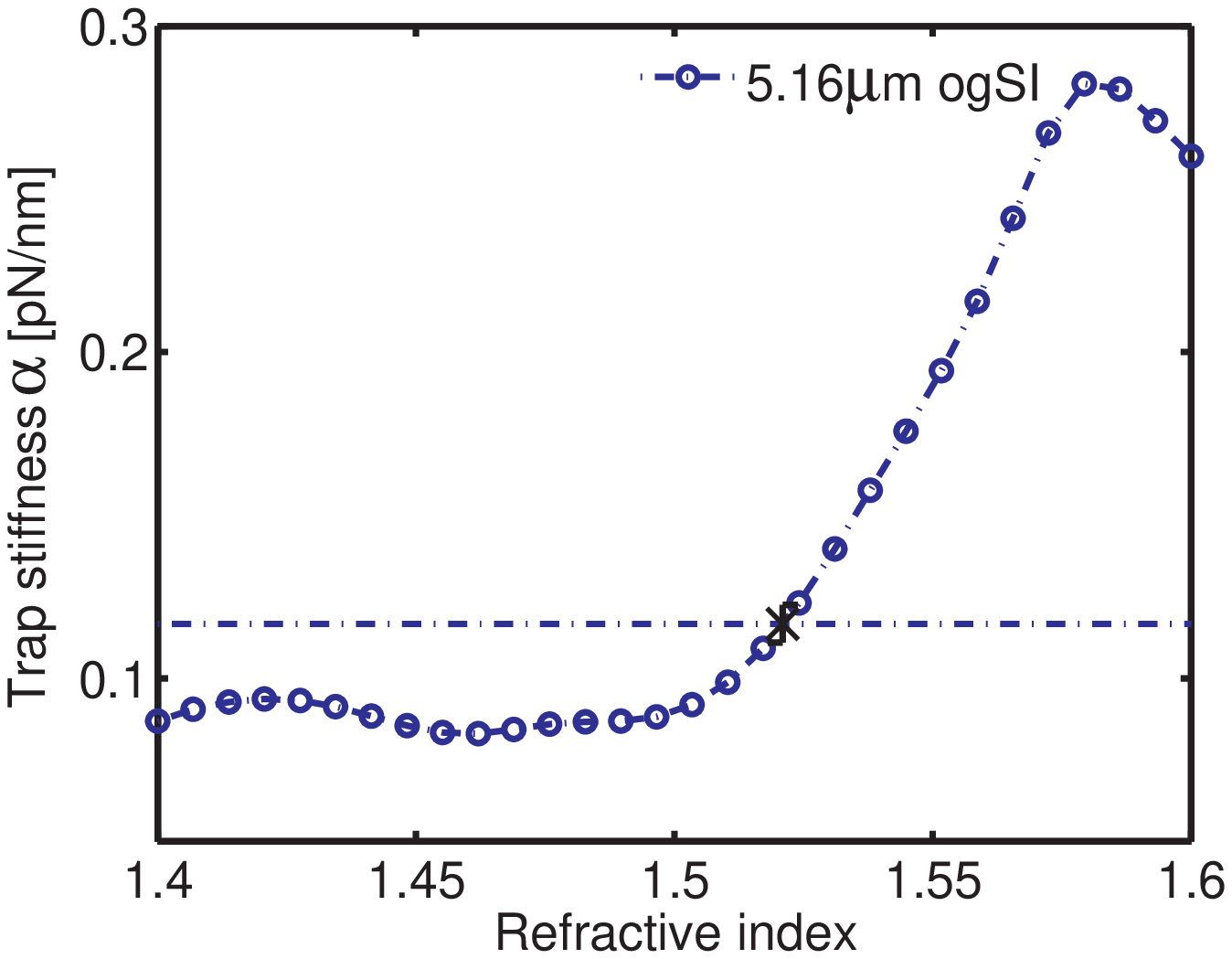}}
\end{tabular}
\caption{Theoretical plots of trap stiffness versus $n_{sp}$  for (a)
2.09 $\mu$m polystyrene (PS), and (b) 5.16 $\mu$m organosilica
microspheres (ogSI).  The intersection of the theoretical curve and the
measured trap stiffness (horizontal line) yields $n_{sp}$.}
\label{FigTrapRI}
\end{figure}

The refractive index of the microspheres was also measured using optical
tweezers refractometry.  This method did not require monodisperse
suspensions since only a single particle was measured at a time.  For
2.09 $\mu$m polystyrene and 5.16 $\mu$m organosilica microspheres, the
power spectrum amplitude of Brownian motion, shown in Fig.
\ref{FigPowerSpec}, was recorded, and the roll-off frequencies
determined.  Using Eq. (\ref{EqCauchy}) and tabulated drag constants,
the trap stiffness was calculated, and is shown in Table
\ref{TabTweezer}.  By using experimentally measured diameters of the
trapped polystyrene and organosilica microspheres we calculated $\alpha$
versus $n_{sp}$ in the framework of the Lorenz-Mie scattering theory,
with the results presented in Fig. \ref{FigTrapRI}.  The intersection of
the theoretical curves and corresponding horizontal lines of the
determined values of $\alpha$ yielded the refractive index of the
trapped microspheres.  For polystyrene microspheres, we obtained a
refractive index of $n_{sp}$ = 1.58 $\pm$ 0.01 at $\lambda$ = 1070 nm. 
This value deviates by a 0.6\% from that given in
literature~\cite{Ma200348}.  For the organosilica microspheres, we
obtained refractive indices of 1.52 $\pm$ 0.01 and 1.51 $\pm$ 0.01 at
589 and 1070 nm respectively, which was in a good agreement with the
respective values of 1.518 and 1.511 obtained by the scattering
spectroscopy method.  For comparison with the scattering spectroscopy
results, the refractive index values obtained by laser tweezers
refractometry are also shown in Fig. \ref{FigDispersion}(a) and (b).\\

\begin{table}
\begin{center}
\caption{Parameters determined by Laser Tweezers Refractometry}
\begin{tabular}{| l | l | l | l | l |}
\hline
Microsphere & $f_o$ (Hz) & $\alpha$ (pN/nm) & $\beta$ ($\times$10$^{-9}$pN/nm)\\ \hline
2.09 $\mu$m PS	& 2714	& 0.31	& 19.72	\\
5.16 $\mu$m ogSI	& 383 &	0.117 &	48.6 \\ \hline
\end{tabular}
\label{TabTweezer}
\end{center}
\end{table}

\section{Conclusions}
The refractive index of organosilica microspheres was determined in the
300--1100 nm wavelength range using laser tweezers refractometry and a
relatively simple scattering spectroscopy technique.  The refractive
index of organosilica has been found to range from 1.60--1.51 over the
wavelength range of 300--1100 nm, which is considerably greater than
that of the organosilica substrate, fused silica.  The development of
the refractometry methods addressed in this paper can be useful for
accurate determination of nano(micro)material refractive index in situ,
whereas the tabulation of refractive index of organosilica microspheres
has important applications to biological screening using flow
cytometry.

\section*{Acknowledgements}
We acknowledge the support of OzNanoLife, which is a project supported
by the International Science Linkage program (CG060027).  This work was
also supported by the Australian Research Council(DP041527 and
FF0455861).


\end{document}